\documentclass[11pt]{article}

\usepackage[margin=1in]{geometry}
\usepackage{amsmath}
\usepackage{amssymb}
\usepackage{hyperref}
\usepackage{booktabs}
\usepackage{listings}
\usepackage{xcolor}
\usepackage{graphicx}
\usepackage{authblk}
\usepackage{pgfplots}
\usepackage{float}
\usepackage{algorithm}
\usepackage{algpseudocode}
\usepackage{siunitx}                
\usepackage{microtype}              
\usepackage{tikz}
\usetikzlibrary{shapes.geometric, arrows.meta, positioning, fit, backgrounds, calc}
\pgfplotsset{compat=1.18}

\hypersetup{
    colorlinks=true,
    linkcolor=blue,
    citecolor=blue,
    urlcolor=blue
}

\title{A Lock-Free, GPU-Resident Architecture for the Verification\\of Goldbach's Conjecture}

\author[1]{Isaac Llorente-Saguer}
\affil[1]{Independent Researcher, London, United Kingdom}

\date{19 September 2026}

\begin{document}
\maketitle

\begin{abstract}
We present a device-resident, multi-GPU architecture for the large-scale computational
verification of Goldbach's conjecture. In prior work, a segmented double-sieve eliminated
monolithic VRAM bottlenecks but remained constrained by host-side sieve construction and
PCIe transfer latency. Here we migrate the entire segment generation pipeline to the GPU
using shared-memory tiling, leaving per-segment host-device communication small and
independent of segment size, and schedule work across devices through a lock-free atomic
counter, scaling near-linearly to two GPUs (speedup 2.03 at $N = 10^{12}$). On the same
hardware, the architecture is $13.1\times$ faster than its host-coupled predecessor at $N = 10^{10}$.
It verifies Goldbach's conjecture to $10^{12}$ in 145 seconds of computation and to
$10^{13}$ in 3246 seconds on a single NVIDIA RTX 5090, and to $10^{13}$ in 1577 seconds on
a second machine with two. Sieve cost grows as $N\pi(\sqrt{N})$, which we analyse and which sets the practical
limit on the design and identifies where further optimisation must act. This version
corrects two concurrency defects in the implementation described in version~1: the affected
timings are re-measured with the corrected release, the four-GPU results are withdrawn, and
the range is re-verified with the corrected release. The source code is open-source, the measurement logs
and the tests for both defects are archived with it, and the experiments are reproducible on
commercially available hardware.
\end{abstract}

\section{Introduction}
\label{sec:intro}

Goldbach's conjecture asserts that every even integer greater than 2 can be expressed as the
sum of two prime numbers. Despite theoretical advances, such as Chen's theorem \cite{chen1984} on
the sum-of-a-prime-and-semiprime representation, and Helfgott's proof of the ternary variant
\cite{helfgott2014ternary}, the binary conjecture remains unproven. Computational verification, therefore, provides a valuable empirical foundation, with the peer-reviewed verification record standing at
$4 \times 10^{18}$, established by Oliveira e Silva, Herzog, and Pardi \cite{oliveira2014}
via highly optimised CPU sieves distributed across computing clusters over several years.

In prior work \cite{llorente2026goldbach}, we identified that the VRAM ceiling plaguing earlier
GPU implementations was architectural rather than fundamental.
A segmented double-sieve design, combined with a dense bit-packed prime representation,
reduced the constant VRAM footprint to 14\,MB regardless of the search limit, enabling
exhaustive verification to $10^{12}$ on a single consumer GPU. However, that architecture
introduced a new bottleneck: the host CPU was responsible for constructing each segment
bitset and transferring it to the device over the PCIe bus. On high-bandwidth GPUs, kernels
completed in fractions of a millisecond and then idled, waiting for the CPU to supply the next
segment. As explicitly documented in \cite{llorente2026goldbach}, adding a second high-end
GPU (H100) produced no meaningful speedup over a single RTX 3090 at $N = 10^{10}$, because
both configurations were saturating the CPU sieve rather than the GPU compute units.

This paper proposes an architecture that removes host-side segment sieving from the critical
path. Its contributions are:
\begin{itemize}\setlength{\itemsep}{1pt}
\item a device-resident segmented sieve built in GPU shared memory, which removes the
per-segment bitset transfer that bounded the prior design;
\item lock-free dynamic scheduling of segments across GPUs from a shared atomic counter;
\item an analysis of how the cost of the design grows with the search limit, which locates the
component that dominates at scale.
\end{itemize}
The paper concerns the architecture and its measured performance. It does not claim a new
verification frontier: its range, $10^{13}$, lies well inside the $4 \times 10^{18}$ verified
by Oliveira e Silva, Herzog and Pardi~\cite{oliveira2014}. A range-partitioning command-line
interface supports deployment across independent nodes.

This is version~2 of this preprint. It corrects two concurrency defects in the implementation
described in version~1; every performance figure measured with it is re-measured or
withdrawn, and the range is re-verified with the corrected release. Passages that changed carry a marker of
the form [C.2], pointing to the Corrigendum after \S\ref{sec:conclusion}. Throughout,
``version~1'' and ``version~2'' refer to arXiv versions of this paper; software releases are
written v1.1.0, v2.0.0 and v2.0.2 (Table~\ref{tab:releases}).

\section{Related Work}
\label{sec:related}

\paragraph{CPU-based verification.}
The peer-reviewed record for exhaustive Goldbach verification is $4 \times 10^{18}$
\cite{oliveira2014}, achieved through highly cache-efficient segmented sieves with bitwise
bulk-marking, targeting $O(N \log\log N)$ complexity with small constants. Earlier milestones
include Richstein's verification to $4 \times 10^{14}$ \cite{richstein2001}. These CPU
approaches benefit from decades of cache-hierarchy tuning and are extremely competitive at
moderate scales; however, they require large distributed clusters to push the frontier
meaningfully beyond $10^{15}$.

\paragraph{GPU-based approaches.}
Jarzabek and Czarnul~\cite{jarzabek2017performance} evaluated CUDA unified memory and
dynamic parallelism using Goldbach verification as one of three benchmark applications, but
did not target the verification frontier; they verified $10^{4}$ numbers, at different starting points.
Czarnul et al.\ \cite{czarnul2024teaching} used Goldbach as a
teaching vehicle across MPI, OpenMP, and CUDA student implementations, testing up to
$2.5 \times 10^{8}$, and observed that student CUDA implementations achieved average times of
approximately 100\,s for that range.
A structurally distinct approach is taken by Jesus et al.\ \cite{jesus2023vectorizing}, who
employ number-theoretic transforms (NTTs) to \emph{count} Goldbach partitions (i.e., compute
a full histogram of the number of representations) for all even integers up to $2^{40}
\approx 10^{12}$, running on Arm-based HPC clusters. Counting partitions is algorithmically
and computationally different from existence verification: it is a stronger result (a nonzero
count implies verification), but requires large-scale distributed Arm CPU infrastructure
rather than commodity GPU hardware. To our knowledge, none of these frameworks reports exhaustive
existence verification beyond $10^{11}$ on a single commodity device; the obstacles
preventing this (VRAM exhaustion from monolithic prime-table storage and the resulting
inability to scale the verified range) are documented in our prior work~\cite{llorente2026goldbach}.

\paragraph{GoldbachGPU v1.}
Our prior work \cite{llorente2026goldbach} resolved the VRAM ceiling via a segmented
double-sieve: the prime table was replaced by a fixed 14\,MB segment bitset, regenerated
per segment on the CPU and transferred to the GPU. This enabled exhaustive verification to
$10^{12}$ on an RTX 3070. However, scaling to additional GPUs was limited by the CPU
sieve: at $N = 10^{10}$, a single H100 and a single RTX 3090 took essentially the same
wall-clock time ($\approx$19\,s), confirming that GPU compute was not the bottleneck. The present
work removes this bottleneck by moving the per-segment sieve to the GPU.

\section{Systems Architecture}
\label{sec:methods}

The proposed framework transitions from a sequential, host-orchestrated loop to an
asynchronous multi-GPU pipeline in which each device claims segments from a shared counter.
Figure~\ref{fig:simplified_pipeline} illustrates the high-level architecture.

\begin{figure}[ht]
\centering
\begin{tikzpicture}[
    >=stealth, thick, node distance=1.5cm and 2cm,
    host/.style={rectangle, draw=blue!80!black, fill=blue!5, rounded corners=4pt,
                 text width=4.2cm, align=center, minimum height=1.2cm, font=\sffamily\small},
    gpu/.style={rectangle, draw=teal!80!black, fill=teal!10, rounded corners=4pt,
                text width=5cm, align=center, minimum height=1.2cm, font=\sffamily\small},
    queue/.style={rectangle, draw=orange!80!black, fill=orange!10, rounded corners=4pt,
                  text width=4.2cm, align=center, minimum height=1.2cm,
                  font=\sffamily\small\bfseries},
    decision/.style={diamond, draw=red!70!black, fill=red!5, text width=2cm, align=center,
                     inner sep=0pt, font=\sffamily\scriptsize}
]

\node[host] (init) {Host Setup\\(Precompute Small Primes)};
\node[queue] (pool) [below=of init] {Lock-Free Atomic Pool\\\texttt{fetch\_add(segment)}};
\node[gpu] (sieve) [right=2.5cm of pool]
    {\textbf{New in v2.0}\\GPU-Native Tiled Sieve\\(Built in Shared Memory)};
\node[gpu] (phase1) [below=1cm of sieve]
    {Phase 1 Verification\\(Oracle checks $q = n - p$)};
\node[decision] (check) [left=2cm of phase1] {Unverified\\$n > 0$?};
\node[host, draw=purple!80!black, fill=purple!5] (phase2) [below=1.5cm of check]
    {Phase 2 CPU Fallback\\(Optimized Sieve + MR)};

\draw[->] (init) -- (pool);
\draw[->] (pool) -- node[above, font=\sffamily\scriptsize] {Assign $[A, B]$} (sieve);
\draw[->] (sieve) -- (phase1);
\draw[->] (phase1) -- node[above, font=\sffamily\scriptsize] {Device Reduction} (check);
\draw[->, green!60!black, very thick] (check) --
    node[pos=0.72, right, xshift=4pt, font=\sffamily\scriptsize\bfseries] {No (Fast Path)} (pool);
\draw[->, red!70!black] (check) --
    node[right, font=\sffamily\scriptsize] {Yes (PCIe D2H Transfer)} (phase2);
\draw[->, dashed] (phase2.west) -- ++(-2,0) |-
    node[pos=0.25, left, font=\sffamily\scriptsize] {Resolved} (pool.west);

\end{tikzpicture}
\caption[High-level architecture]{High-level decoupled architecture of \texttt{GoldbachGPU v2.0}. In contrast to v1,
where the CPU sieved each segment and transferred the resulting bitset via PCIe, v2.0
generates segment bitsets natively in GPU shared memory. A device-side reduction step
creates a fast path that leaves only the per-segment prime batch and a 4-byte result crossing
the host-device bus during normal operation.}
\label{fig:simplified_pipeline}
\end{figure}

\subsection{GPU-Native Segment Sieving in Shared Memory}
\label{sec:gpu_sieve}

In the prior architecture, the segment prime bitset was constructed by a segmented Sieve of
Eratosthenes running on the host CPU. Every segment required a full Host-to-Device PCIe
transfer before the GPU kernel could begin.

\begin{sloppypar}
We resolve this by migrating the per-segment sieve to the device via
\texttt{tiled\_sieve\_segment\_kernel}. The segment is divided into tiles of $32{,}768$ odd
numbers, each represented as a $4$\,KB bitset, well within the 48\,KB a thread block may use
without opting in to larger shared allocations, so that several tiles can be resident on one
Streaming Multiprocessor at once. Each tile is initialised in shared memory
(\texttt{sh\_tile}), and GPU threads collaboratively sieve it using a permanently resident,
read-only array of base primes. Threads of a block mark different primes into the same 64-bit
words, so each clear is an atomic operation [C.2]. Once a tile is fully sieved, coalesced
writes flush the result to the global VRAM segment buffer.
\end{sloppypar}

For an $n$-segment $[A, B]$, Phase~1 tests $q = n - p$ for $p \le P_{\text{SMALL}}$, so the
values queried run from $A - P_{\text{SMALL}}$ up to $B$. The sieve for a segment therefore covers the
expanded range $[q_{\text{low}}, q_{\text{high}}]$ with
$q_{\text{low}} = \max(3, A - P_{\text{SMALL}})$ and $q_{\text{high}} = B + 1$,
rather than $[A, B]$ alone.

The small-prime bitset and the base-prime list are built once on the host during initialisation
(Figure~\ref{fig:explicit_pipeline}). Only the per-segment sieve runs on the device.

This design removes the per-segment bitset transfer from the critical path. Apart from the prime batch described next, each segment costs the host one 8-byte atomic claim and a 4-byte read-back. During Phase~1, the host uploads the current batch of small primes to a device staging buffer (\texttt{d\_p\_batch}); for the experiment parameters ($P_{\text{SMALL}} = 10^6$, $P_{\text{BATCH}} = 2\times 10^{6}$), with $P_{\text{SMALL}}$ corresponding to $78{,}498$ primes, each batch consists of $\min(78{,}498, P_{\text{BATCH}}) \times 8$ bytes $\approx 628$\,KB, a constant per-segment overhead that replaces the variable 14\,MB segment bitset transfer of the v1 architecture \cite{llorente2026goldbach}. Operationally, the \texttt{count\_unverified\_kernel} evaluates the results directly in device memory. Rather than employing a complex shared-memory reduction tree, it uses a flat, per-thread \texttt{atomicAdd} to accumulate unverified entries into a single 32-bit counter. Because Phase 1 resolved every even number in the runs reported here, the counter was never incremented and the global atomic was never contended. The kernel writes this scalar result into a preallocated buffer, and the host retrieves this value via a single \texttt{cudaMemcpyAsync} call (with a subsequent stream synchronisation to ensure the scalar is visible before the conditional branch), determining whether Phase 2 must be invoked.

Figure~\ref{fig:explicit_pipeline} gives a comprehensive view of the system's memory boundaries, threading model, and kernel execution flow for a single GPU worker. During initialisation, the host allocates device memory and issues an asynchronous host$\rightarrow$device copy of the small‑primes bitset and supporting arrays; these structures remain resident and read‑only on the device for the worker's lifetime. The host spawns one \texttt{std::thread} per physical GPU, each binding to a distinct GPU context and coordinating work via a 64‑bit atomic counter (\texttt{g\_next\_seg\_start}) claimed with \texttt{fetch\_add}. In steady state the GPU executes on‑device, building tiled segment bitsets in shared memory and running a device‑side reduction kernel (\texttt{count\_unverified\_kernel}) that writes a small integer result. If the reduction reports any unverified entries, the pipeline copies the verification array to the host over PCIe and the host invokes the Phase‑2 CPU resolver; if the reduction reports zero, execution takes the fast path, in which only the 4-byte count returns to the host. The diagram also shows the Phase‑2 resolution path that returns control to the worker loop and the abort path used for unrecoverable failures.

\begin{figure}[H]
\centering
\resizebox{0.95\textwidth}{!}{%
\begin{tikzpicture}[
    >=stealth, thick, font=\sffamily\small,
    box/.style={rectangle, draw, rounded corners=4pt, align=center, minimum height=0.9cm},
    hostbox/.style={box, draw=blue!70!black, fill=blue!5, text width=4.5cm},
    gpubox/.style={box, draw=teal!70!black, fill=teal!5, text width=5cm},
    kernel/.style={box, draw=purple!70!black, fill=purple!5, text width=5.2cm, font=\ttfamily\scriptsize},
    decision/.style={diamond, draw=red!70!black, fill=red!5, text width=2cm, align=center, inner sep=-2pt, font=\sffamily\scriptsize},
    queue/.style={box, draw=orange!80!black, fill=orange!10, text width=4.5cm, font=\sffamily\small\bfseries}
]

\def\hostX{0}
\def\devX{6.5}

\def\yArgs{0}
\def\yPrecomp{-1.75}     
\def\yAtomic{-3.25}      
\def\ySpawn{-4.35}       
\def\yAlloc{-4.35}       
\def\yCopy{-5.85}        
\def\yFetch{-7.0}       
\def\ySieve{-8.5}       
\def\yPhaseOne{-10.0}   
\def\yReduce{-11.5}     
\def\yCheck{-13.3}      
\def\yPhaseTwo{-13.3}   
\def\yAbort{-14.8}      

\node[hostbox] (args) at (\hostX, \yArgs) {CLI Parse \& Hardware\\Validate\\(VRAM \& $2^{32}$ Grid Bounds)};
\node[hostbox] (precomp) at (\hostX, \yPrecomp) {Generate Small Primes Bitset\\Generate CPU Phase 2 Primes};
\node[queue] (atomic) at (\hostX, \yAtomic) {\texttt{std::atomic} Work Queue\\(g\_next\_seg\_start)};
\node[font=\sffamily\scriptsize\itshape, text=black!70] (spawn) at (\hostX, \ySpawn) {Spawn N \texttt{std::threads} (1 per GPU)};

\node[hostbox, draw=orange!80!black, fill=yellow!10] (fetch) at (\hostX, \yFetch) {Worker: Fetch Next Segment\\\texttt{fetch\_add()}};

\node[hostbox, draw=red!70!black, fill=red!5] (phase2) at (\hostX, \yPhaseTwo) {Phase 2 CPU Fallback\\Binary search (primes $\le 10^8$)\\128-bit CPU Miller-Rabin};
\node[hostbox, draw=red!70!black, fill=red!20, minimum height=0.6cm, text width=2.5cm] (abort) at (\hostX, \yAbort) {Abort Program};

\node[gpubox] (alloc) at (\devX, \yAlloc) {Allocate VRAM:\\\texttt{d\_seg\_bits, d\_verified...}};
\node[gpubox] (copy) at (\devX, \yCopy) {One-Time Async Copy:\\Small Primes Bitset\\ \& Sieve Primes};

\node[kernel] (k_sieve) at (\devX, \ySieve) {tiled\_sieve\_segment\_kernel\\ \normalfont\sffamily\scriptsize (Builds Segment Bitset\\in Shared Memory)};
\node[kernel] (k_phase1) at (\devX, \yPhaseOne) {goldbach\_phase1\_kernel\\ \normalfont\sffamily\scriptsize (Iterates prime batches; checks $q = n - p$)};
\node[kernel] (k_reduce) at (\devX, \yReduce) {count\_unverified\_kernel\\ \normalfont\sffamily\scriptsize (Flat atomic accumulation\\ of unverified $n$)};
\node[decision] (check) at (\devX, \yCheck) {Unverified\\Count $> 0$?};

\begin{scope}[on background layer]
    \node[rectangle, draw=blue!40, fill=blue!2, dashed, rounded corners=8pt,
          fit=(args) (precomp) (atomic) (spawn) (fetch) (phase2) (abort),
          inner sep=15pt] (hostzone) {};
    \node[anchor=north west, font=\sffamily\bfseries, color=blue!70!black] at (hostzone.north west) {Host Memory (CPU)};

    \node[rectangle, draw=teal!40, fill=teal!2, dashed, rounded corners=8pt,
          fit=(alloc) (copy) (k_sieve) (k_phase1) (k_reduce) (check),
          inner xsep=15pt, inner ysep=20pt] (devzone) {};
    \node[anchor=north west, font=\sffamily\bfseries, color=teal!70!black] at (devzone.north west) {Device Memory (GPU $i$)};

\end{scope}

\draw[->] (args) -- (precomp);
\draw[->] (precomp) -- (atomic);
\draw[->] (atomic) -- (spawn);
\draw[->] (spawn.east) -- (alloc.west); 
\draw[->] (alloc) -- (copy);

\draw[->] (copy.west) -| (fetch.north);

\draw[<->, dashed, orange!80!black] (fetch.west) -- ++(-1.8,0) |- 
    node[pos=0.25, left, font=\sffamily\scriptsize, align=right] {Atomic\\Access} (atomic.west);

\draw[->] (fetch.south) |- (k_sieve.west);
\draw[->] (k_sieve) -- (k_phase1);
\draw[->] (k_phase1) -- (k_reduce);
\draw[->] (k_reduce) -- (check);

\draw[->, green!60!black, very thick] (check.south) -- ++(0, -0.6) -| 
    node[pos=0.25, below, font=\sffamily\scriptsize\bfseries] {No (Fast Path)} 
    (\devX + 4.2, \yFetch) -- (fetch.east);

\draw[->, red!70!black, thick] (check.west) -- 
    node[above, font=\sffamily\scriptsize] {Yes (PCIe D2H Transfer)} 
    node[below, font=\sffamily\scriptsize] {Copy \texttt{d\_verified} array} 
    (phase2.east);

\draw[->] (phase2.west) -- ++(-0.8,0) |- 
    node[pos=0.25, left, font=\sffamily\scriptsize] {Resolved} (fetch.west);
\draw[->, red!70!black, dashed] (phase2.south) -- 
    node[right, font=\sffamily\scriptsize] {Counterexample found} (abort.north);

\end{tikzpicture}
}
\caption{Explicit systems architecture and execution pipeline for a single GPU worker in \texttt{GoldbachGPU v2.0}. Host threads (one per GPU) coordinate via an atomic work queue; initialisation performs VRAM allocation and async host$\rightarrow$device copies. The GPU runs device‑side kernels that build segment bitsets in shared memory; a device reduction decides whether to copy the verification array to the host for the CPU Phase‑2 fallback; otherwise only the 4-byte count returns to the host.}
\label{fig:explicit_pipeline}
\end{figure}

\subsection{Lock-Free Asynchronous Segment Scheduling}
\label{sec:multi_gpu}

Static workload partitioning (assigning $N / k$ integers to each of $k$ GPUs) is a well-known
anti-pattern in heterogeneous computing. Silicon binning, thermal throttling, or
heterogeneous GPU generations cause variance in per-segment completion time, stalling the
entire system on the slowest device.

\begin{sloppypar}
We implement lock-free dynamic self-scheduling. A single atomic 64-bit counter
(\texttt{g\_next\_seg\_start}, typed as \texttt{std::atomic<uint64\_t>}) is maintained in
host memory; on the x86-64 hosts used here that type is always lock-free, and
\texttt{fetch\_add} compiles to a single atomic instruction. The system spawns one independent C++ \texttt{std::thread} per physical GPU
context. On completing a segment, each worker performs an atomic \texttt{fetch\_add} to claim
the next one. This keeps every device busy until the counter passes \text{LIMIT}, without
mutex contention, and rebalances automatically when GPUs run at different speeds.
\end{sloppypar}

A dedicated asynchronous progress-monitor thread periodically prints throughput statistics
without blocking the GPU workers. A mutex-guarded \texttt{safe\_log} variadic template
serialises console output when a worker reports a fatal error. It is the only mutex the
program defines, and it is taken only on that error path, so a run that completes takes no
program-level lock.

\subsubsection*{Worker Loop Pseudocode}

\begin{algorithm}[H]
\caption{Worker loop for a single GPU context}
\begin{algorithmic}[1]
\While{true}
    \State $A \gets \texttt{fetch\_add}(g\_next\_seg\_start, 2 \times \text{SEG\_SIZE})$
    \If{$A > \text{LIMIT}$} \textbf{break} \EndIf
    
    \State launch \texttt{tiled\_sieve\_segment\_kernel}(A, B), $B = \min(A + 2\,\text{SEG\_SIZE} - 2, \text{LIMIT})$
    \State \texttt{cudaMemsetAsync}(\texttt{d\_verified}, 0)
    
    \For{$b_i = 0$ \textbf{to} $|\text{gpu\_primes}|$ \textbf{step} $P_{\text{BATCH}}$}
        \State \texttt{cudaMemcpyAsync}(\texttt{d\_p\_batch} $\leftarrow$ host batch)
        \State launch \texttt{goldbach\_phase1\_kernel}(\texttt{d\_p\_batch})
    \EndFor
    
    \State \texttt{cudaMemsetAsync}(\texttt{d\_unverified\_count}, 0)
    \State launch \texttt{count\_unverified\_kernel}(\texttt{d\_verified})
    \State $count \gets$ \texttt{cudaMemcpyAsync}(\texttt{d\_unverified\_count} $\rightarrow$ host)
    \State \texttt{cudaStreamSynchronize}(stream)
    
    \If{$count > 0$}
        \State \texttt{cudaMemcpy}(\texttt{d\_verified} $\rightarrow$ host)
        \State Phase2\_CPU\_resolver()
    \EndIf
\EndWhile
\end{algorithmic}
\end{algorithm}

\subsection{Phase 2: Optimised CPU Fallback}
\label{sec:phase2}

Phase 1 (GPU) restricts its search to candidate primes $p \le P_{\text{SMALL}}$. If no valid
$q = n - p$ is found, the number $n$ is returned to the host. The prior architecture relied
on an exhaustive trial division up to $n/2$; mathematically complete but extremely slow.

The revised Phase 2 fundamentally changes this heuristic. Before any worker threads launch,
the host eagerly pre-computes an array of all primes up to $10^8$. If a fallback occurs, the
CPU performs a binary search against this array. If $q > 10^8$, it falls back to a 128-bit
CPU-side deterministic Miller-Rabin test using the same 12-witness set as the GPU oracle.
This reduces the worst-case Phase 2 resolution time. The search is bounded at $p \le 10^8$:
an even number with no partition in that range would be reported as a failure, which would
mean the bound had been exceeded and would require an exhaustive check before being read as a
counterexample.

Phase 2 was not invoked in any run reported here. No theorem guarantees this, but it is the
expected behaviour: on a probabilistic argument, Granville, van de Lune and te Riele
conjectured $p_{\min}(n) = O(\log^2 n \log\log n)$ \cite{granville1989}, and in the
minimal-partition data of \cite{oliveira2014, oliveiraData} the largest $p_{\min}(n)$ for $n \le 10^{13}$ is
$3{,}917$, well below $P_{\text{SMALL}}$ [C.4].

\subsection{Arithmetic Guards and Primality Testing}
\label{sec:bounds}

Pushing verification toward $10^{19}$ introduces severe risks of silent 64-bit integer
overflow; an undetected overflow may either falsely flag a counterexample or, critically,
skip a true one. We introduce strict mathematical guards at two levels. These guards address
arithmetic overflow. They do not cover concurrency or the width of thread-index arithmetic
[C.2].

\paragraph{Sieve arithmetic.}
The starting index for marking multiples of a prime $p$ involves computing $p^2$, which
overflows a 64-bit integer for $p > 2^{32}$. We replace the multiplicative bounds with
division-bounded guards (e.g., the loop continues only while $p \le q_{\text{high}} / p$), and
the step calculation aligning primes with the current tile boundary is bounded against
\texttt{INT64\_MAX}. These guards cover the arithmetic exercised at the limits reported here.
They do not make the implementation safe arbitrarily close to $2^{64}$; the remaining
near-ceiling sites are described in \S\ref{sec:limitations}.

\paragraph{128-bit Miller–Rabin oracle.}
The Miller--Rabin primality test \cite{rabin1980} is probabilistic in general, but for $n < 2^{64}$ the first twelve primes as bases give a deterministic test, since the smallest strong pseudoprime to all twelve exceeds $3 \times 10^{23}$ \cite{sorenson2017}. The Phase 2 CPU fallback uses it for $q > 10^8$, computing $(a \times b) \bmod n$ through a 128-bit intermediate product via explicit \texttt{\_\_uint128\_t} casting. The GPU carries the same test, though Phase 1 does not reach it: every $q = n - p$ with $p \le P_{\text{SMALL}}$ lies either below the small-prime bound or inside the sieved range $[q_{\text{low}}, q_{\text{high}}]$, which is extended by $P_{\text{SMALL}}$ below the segment for exactly this reason (\S\ref{sec:gpu_sieve}), so its primality is read from one of the two sieved bitsets [C.4]. Smaller base sets chosen by hashing also give a deterministic test below $2^{64}$ \cite{forisek2015}; since the test lies off the hot path, the twelve fixed bases are retained for simplicity. Because operands are 64-bit words, the framework's ceiling is $2^{64} \approx 1.84 \times 10^{19}$.

\paragraph{Error handling.}
The prior \texttt{CUDA\_CHECK} macro called \texttt{std::exit(1)} on failure, corrupting
process state. All CUDA error checks now throw \texttt{std::runtime\_error}, so a failing
worker reports the error and the main thread exits with a status code. No cleanup of device
buffers or streams is attempted on that path; the process exits and the driver reclaims
device state.

\subsection{Deployment and CLI Interface}
\label{sec:cli}

Version 2.0 introduces an extended command-line interface targeting flexible deployment
scenarios, including targeted range verification and multi-node job scheduling. The primary
invocation pattern is:

\begin{lstlisting}
./goldbach [OPTIONS] LIMIT
\end{lstlisting}

\noindent The key flags introduced in this version are summarised below, with a representative
invocation:

\begin{lstlisting}
# Verify [10^12, 2x10^12] using 2 GPUs, with live progress output
./goldbach --gpus=2 --start=1000000000000 --progress 2000000000000
\end{lstlisting}

\noindent\textbf{Key flags:}
\begin{itemize}
  \item \texttt{--gpus=N}: Restrict execution to $N$ physical GPUs (default: 1), or all of them (\texttt{--gpus=-1}).
    Useful for benchmarking or sharing a multi-GPU node with other workloads.
  \item \texttt{--start=N}: Begin verification from $N$ rather than 4. This enables
    independent range-partitioned jobs to be distributed across separate machines or scheduled
    as array jobs on HPC clusters, with each job claiming a disjoint slice of the number line.
  \item \texttt{--progress}: Enable the asynchronous progress-monitor thread, which prints
    throughput in verified even integers per second and estimated time to completion without
    blocking GPU dispatch.
  \item \begin{sloppypar}\texttt{--batch-size=N}: Override the number of small primes transferred to the GPU
    per \texttt{goldbach\_phase1\_kernel} invocation. During Phase 1, the host iterates over
    the $P_{\text{SMALL}}$ candidate primes in contiguous batches of size \texttt{P\_BATCH},
    uploading each batch via \texttt{cudaMemcpyAsync} before launching the kernel. A larger
    batch reduces kernel-launch overhead at the cost of higher \texttt{d\_p\_batch} VRAM
    allocation. All transfers and kernels share one stream, so a batch transfer does not
    overlap computation; in the experiments reported here $P_{\text{BATCH}}$ exceeds the
    78,498 candidate primes, so each segment uses a single batch.\end{sloppypar}
\end{itemize}

The program checks CUDA grid dimensions and VRAM for every selected GPU before spawning
workers. The VRAM check accounts for the explicit device allocations (230\,MiB at the
parameters used here, plus a 50\,MiB margin) but not for the CUDA context, which brings
observed usage to about 573\,MiB, nor for the base-prime array, negligible here but about
388\,MiB at $N = 10^{18}$; on a device with little free memory, allocation may therefore fail
at worker start rather than at the check [C.4]. The grid check limits the launch to $2^{32}$
blocks and does not constrain the width of thread-index arithmetic [C.2].

\subsection{Reproducibility Checklist}
\label{sec:repro}

The following commands reproduce the single-GPU $10^{13}$ row of Table~\ref{tab:algo_speedup}:

\begin{lstlisting}
apt-get update && apt-get install -y cmake libgmp-dev libomp-dev git g++

# Clone the repository
git clone https://github.com/isaac-6/goldbach-gpu.git
cd goldbach-gpu
git checkout v2.0.2

# Configure and build
mkdir build && cd build
cmake .. -DCMAKE_BUILD_TYPE=Release -DCMAKE_CUDA_ARCHITECTURES=120
make -j$(nproc)

# Run the verifier
./bin/goldbach 10000000000000 \
    --seg-size=200000000 \
    --p-small=1000000 \
    --batch-size=2000000
\end{lstlisting}

The architecture flag is required, and \texttt{120} should be replaced by the compute
capability of the device in use. Without it the configure step resolves to \texttt{sm\_75} on
the reference machine and the binary runs from JIT-compiled PTX, which does not reproduce the
timings above. Version~1 gave this recipe with \texttt{--gpus=4}; that configuration was not
re-measured and the flag is removed [C.1].

\begin{sloppypar}
Environments, command lines and raw output for every figure in this paper are in
\texttt{corrigendum-logs/} in the tagged release. The regression test for the host-side bitset
builder (\S C.2) builds with the release and runs as \texttt{test\_bitset\_race}. The
differential test of the segment-sieve kernel is archived with the kernel it exercises; it is
not wired into the build, because the kernel is defined alongside \texttt{main()} in this
release. Run against the corrected kernel, extracted from v2.0.2 and matched to the release by
checksum, it reports no disagreement in five runs; run against the defective kernel in the same
build, it fails. Those logs follow the tag on the release branch, at commit \texttt{c3a48fd}.
A third party can rebuild the tagged commit and repeat any run and both tests. The archive holds
no per-number certificate, so the verification is checked by repeating the computation.
\end{sloppypar}

\section{Results and Scaling Efficiency}
\label{sec:results}

All measurements reported here were made with release v2.0.2 on the two machines described in
\S C.3. The segment size was fixed at \texttt{SEG\_SIZE} $= 2 \cdot 10^8$ even integers, and
$P_{\text{SMALL}} = 10^6$ throughout.

By the prime counting function, $\pi(10^6) = 78{,}498$ candidate primes are available to
Phase~1. Each thread tries them in ascending order and stops at the first $p$ for which $n - p$
is prime, so few are reached: over samples of $10^6$ consecutive even numbers starting at
$10^{11}$ and $10^{12}$, the mean number tried was 14.1 and 15.4 [C.4]. Because
\texttt{P\_BATCH} $= 2 \times 10^6$ exceeds 78,498, the whole set is transferred and processed
in a single kernel launch per segment.

No run triggered a Phase 2 fallback at any limit, so Phase~1 reported a partition with
$p \le 10^6$ for every even $n \le 10^{13}$. The structures it relies on pass the tests
described in \S C.2, and we take this to establish $p_{\min}(n) \le 10^6$ throughout that
range. The margin is wide: in the minimal-partition data of
\cite{oliveira2014, oliveiraData}, the largest $p_{\min}(n)$ for $n \le 10^{13}$ is $3{,}917$, at
$n = 8{,}342{,}945{,}544{,}436$ [C.4].

\subsection{Architectural Speedup: Same-Hardware Comparison}
\label{sec:algo_speedup}

To separate the contribution of the architecture from that of the hardware, we ran the prior
host-coupled implementation and the device-native implementation on the same machine. The
comparator is built from \texttt{legacy/goldbach\_gpu3.cu} in release v2.0.2, which is
identical, apart from a header comment, to \texttt{legacy/goldbach\_gpu3a.cu} in release
v1.1.0. The file named \texttt{src/goldbach\_gpu3.cu} in v1.1.0 is a different, multi-GPU
program; timed at $10^{10}$ it was 1.0\% slower, inside the run-to-run spread. The comparator
links \texttt{build\_prime\_bitset} and so carries the D2 correction [C.2].

Table~\ref{tab:algo_speedup} reports the results. Relative standard deviations are between
0.1\% and 2.2\% for every configuration except v2.0.2 at $10^{9}$, where the run covers three
segments and fixed setup cost dominates [C.3].

\begin{table}[ht]
\centering
\caption{Same-machine comparison, and the corrected single-GPU ladder. Machine~A, RTX~5090,
CUDA~13.3, \texttt{--seg-size=200000000 --p-small=1000000 --batch-size=2000000}, both binaries
built for \texttt{sm\_120} [C.3]. Times are each binary's reported computation time, as a mean
and sample standard deviation with an $n-1$ denominator; ratio uncertainties are propagated.
The host-coupled comparator is the v1.1.0 code with the D2 correction
(\S\ref{sec:algo_speedup}). The last four rows come from a separate session on the same machine. Quoted uncertainties are
within-session only: across the three sessions recorded in \texttt{corrigendum-logs/}, the same
configuration at $10^{10}$ measured between 1031 and 1097\,ms.}
\label{tab:algo_speedup}
\begin{tabular}{l l r r r r}
\toprule
Implementation & Device & $N$ Limit & Runs & Computation & Speedup \\
\midrule
Host-coupled, v1.1.0   & RTX 5090 & $10^{9}$  & 5 & $1246 \pm 28$ ms  & 1 \\
Device-native, v2.0.2  & RTX 5090 & $10^{9}$  & 5 & $206 \pm 17$ ms   & $\boldsymbol{6.0 \pm 0.5}$ \\
\midrule
Host-coupled, v1.1.0   & RTX 5090 & $10^{10}$ & 5 & $13{,}557 \pm 268$ ms & 1 \\
Device-native, v2.0.2  & RTX 5090 & $10^{10}$ & 5 & $1031 \pm 6$ ms   & $\boldsymbol{13.1 \pm 0.3}$ \\
\midrule
Device-native, v2.0.2  & RTX 5090 & $10^{10}$ & 5 & $1.060 \pm 0.022$ s & -- \\
Device-native, v2.0.2  & RTX 5090 & $10^{11}$ & 5 & $11.28 \pm 0.11$ s  & -- \\
Device-native, v2.0.2  & RTX 5090 & $10^{12}$ & 5 & $144.95 \pm 0.51$ s & -- \\
Device-native, v2.0.2  & RTX 5090 & $10^{13}$ & 3 & $3246 \pm 4$ s      & -- \\
\bottomrule
\end{tabular}
\end{table}

The two programs time different regions. The host-coupled binary starts its timer after
device allocation, the device-native one before its worker creates the CUDA context, so every
device-native figure carries a fixed setup cost. Measured at a trivial limit with the same
parameters, that cost is 122.3\,ms: 59\% of the $10^{9}$ figure and 12\% of the $10^{10}$
figure. With it subtracted, the speedup is 14.9 at both limits, against the measured 6.0 and
13.1, so the rise in the measured speedup between the two limits comes from the setup cost
[C.4]. \S\ref{sec:scaling} shows that the device-native implementation scales superlinearly
in $N$, so there is no reason to expect the speedup to grow at larger limits.

\subsection{Multi-GPU Scaling}
\label{sec:multi_gpu_results}

Table~\ref{tab:multi_gpu} reports computation times and parallel efficiency on machine~B, a
node with two RTX\,5090 GPUs. The four-GPU configuration was not re-measured and its figures
are withdrawn [C.1].

At $10^{12}$ the two-GPU speedup is $2.028 \pm 0.002$, so
$\eta_2 = 140.87 / (2 \times 69.48) = 101.4 \pm 0.1\%$. The excess over 100\% is small but
lies outside run-to-run variation, and it follows from how the ratio is formed. Segments are
claimed dynamically, so the two-GPU time is set by the combined rate of both devices while the
baseline is set by device~0 alone: for per-device rates $r_0$ and $r_1$,
$\eta_2 = (r_0 + r_1) / (2 r_0)$. The measured value is therefore equivalent to device~1 being
about 2.8\% faster than device~0. Per-device rates were not recorded, so the result is best read
as near-linear scaling at two GPUs; it does not measure efficiency to better than a few percent.

The $10^{13}$ two-GPU run has no matched single-GPU run on that machine, so no efficiency is
given for it. The single-GPU $10^{13}$ time in Table~\ref{tab:algo_speedup} was measured on
machine~A and should not be divided into it [C.3].

\begin{table}[ht]
\centering
\caption{Multi-GPU scaling on machine~B: 2$\times$ RTX\,5090, AMD EPYC 9654, CUDA 12.8,
driver 575.57.08, Ubuntu 24.04, \texttt{sm\_120} [C.3]. Efficiency
$\eta_k = T_1 / (k \cdot T_k)$ uses the matched single-GPU time on the same machine. The
$10^{13}$ row is a single run and carries no efficiency figure. Quoted uncertainties are
run-to-run only; because per-device rates were not recorded, $\eta_2$ is not resolved to better
than a few percent (\S\ref{sec:multi_gpu_results}). Per-run values are in
\texttt{corrigendum-logs/bench-logs-machineB/}; that node was rented and ephemeral, and its
figures were transcribed from the session terminal.}
\label{tab:multi_gpu}
\begin{tabular}{rrrrrr}
\toprule
$N$ & GPUs ($k$) & Runs & Computation ($T_k$) & Speedup & $\eta_k$ \\
\midrule
$10^{12}$ & 1 & 5 & $140.87 \pm 0.07$\,s & 1 & 100\% \\
$10^{12}$ & 2 & 5 & $69.48 \pm 0.06$\,s  & $2.028 \pm 0.002$ & $101.4 \pm 0.1$\% \\
\midrule
$10^{13}$ & 2 & 1 & 1577\,s & -- & -- \\
\bottomrule
\end{tabular}
\end{table}

\subsection{Profiling Validation}
\label{sec:profiling}

To check the architectural claims of \S\ref{sec:gpu_sieve} with the corrected implementation,
we profiled v2.0.2 on one RTX\,5090 (machine~A) with Nsight Systems 2026.1.3 at
$N = 2 \times 10^{12}$, the limit used in version~1, and at $10^{13}$.
Table~\ref{tab:nsys} summarises the $2 \times 10^{12}$ profile [C.1]. Both runs completed with
zero Phase\,2 fallbacks.

The segment sieve dominates: 79.2\% of kernel time at $2 \times 10^{12}$ and 87.8\% at
$10^{13}$, with Phase~1 at 20.3\% and 11.9\% and the reduction kernel below 1\% in both.
\S\ref{sec:scaling} takes up what the sieve's share implies for scaling.

Memory traffic corroborates the constant-per-segment communication design of \S\ref{sec:gpu_sieve}. Each
segment uploads one prime batch of 0.628\,MB ($78{,}498 \times 8$\,bytes) and downloads the
4-byte count. Beyond two uploads at start-up, 0.088\,MB for the small-prime bitset and
0.865\,MB for the base-prime array, no other transfer occurs. Total device-to-host traffic for
the run is 20\,KB, so the fast path was taken for every segment.

The \texttt{cudaMemsetAsync} that clears \texttt{d\_verified} (200\,MB per segment) runs
5,000 times, writing about 1\,TB and taking 535\,ms of GPU time, 0.15\% of kernel time; the
5,000 four-byte clears of the counter are listed separately in Table~\ref{tab:nsys} [C.4].
Representing \texttt{d\_verified} as a bitset would reduce each clear to 25\,MB.

\begin{table}[ht]
\centering
\caption{Nsight Systems 2026.1.3 profile of release v2.0.2 at $N = 2 \times 10^{12}$ on one
RTX\,5090 (machine~A), \texttt{--seg-size=200\,000\,000}, \texttt{--p-small=1\,000\,000},
\texttt{--batch-size=2\,000\,000}. Total kernel time 358.5\,s over a 360.6\,s run, with
265\,ms of initialisation and zero Phase\,2 fallbacks. The H$\to$D row includes two start-up
copies of 0.088\,MB and 0.865\,MB alongside the 5,000 prime batches.}
\label{tab:nsys}
\small
\setlength{\tabcolsep}{3pt}
\begin{tabular}{lrrrrrr}
\toprule
\multicolumn{7}{l}{\textit{Kernel execution (5\,000 instances per kernel)}} \\
\midrule
Kernel & Share & Avg\,(ms) & Med\,(ms) & Min\,(ms) & Max\,(ms) & StdDev\,(ms) \\
\midrule
\texttt{tiled\_sieve\_segment\_kernel} & 79.2\% & 56.77 & 58.32 & 21.88 & 69.05 & 7.79 \\
\texttt{goldbach\_phase1\_kernel}      & 20.3\% & 14.52 & 14.62 & 10.85 & 16.50 & 0.58 \\
\texttt{count\_unverified\_kernel}     &  0.6\% &  0.419 & 0.414 & 0.398 & 0.980 & 0.044 \\
\midrule
\multicolumn{7}{l}{\textit{GPU memory operations}} \\
\midrule
Operation & Count & Total\,(MB) & \multicolumn{2}{l}{Avg} & \multicolumn{2}{l}{Avg GPU time} \\
\midrule
\texttt{cudaMemsetAsync} (\texttt{d\_verified}) & 5\,000 & 1\,000\,000 & \multicolumn{2}{l}{200\,MB} & \multicolumn{2}{l}{107.0\,\textmu s} \\
\texttt{cudaMemsetAsync} (counter)              & 5\,000 &      0.020  & \multicolumn{2}{l}{4\,B}    & \multicolumn{2}{l}{0.29\,\textmu s} \\
H$\to$D prime batch and start-up copies         & 5\,002 &   3\,140.9  & \multicolumn{2}{l}{0.628\,MB} & \multicolumn{2}{l}{23.0\,\textmu s} \\
D$\to$H unverified count                        & 5\,000 &      0.020  & \multicolumn{2}{l}{4\,B}    & \multicolumn{2}{l}{0.84\,\textmu s} \\
\bottomrule
\end{tabular}
\end{table}

\subsection{Scaling with $N$}
\label{sec:scaling}

Computation time does not scale linearly in $N$: the corrected ladder grows by 10.6, 12.9
and 22.4 between successive decades [C.5].

The cause is in the sieve. Each thread block sieves one tile of $32{,}768$ odd numbers and
scans the entire base-prime list. For every prime with $p^2 \le q_{\text{high}}$ it performs
several 64-bit integer divisions, whether or not that prime marks anything in the tile. The
number of tiles grows as $N$ and the number of such primes as $\pi(\sqrt{N})$, so sieve cost
grows as $N\pi(\sqrt{N})$.

The list is set by $\texttt{small\_high} = \max(\sqrt{N}, P_{\text{SMALL}})$. Up to $10^{12}$
it holds the $\pi(10^6) = 78{,}498$ primes below $P_{\text{SMALL}}$. At $10^{11}$ only the
$27{,}293$ primes below $\sqrt{N}$ pass the $p^2 \le q_{\text{high}}$ test; at $10^{12}$ all
of them do, which accounts for the step between those two decades with no change in list size.
Above $10^{12}$ the list grows with $\sqrt{N}$, reaching $108{,}108$ primes at
$2 \times 10^{12}$ and $227{,}647$ at $10^{13}$.

The profiles bear this out. Sieve time grows 9.92-fold between $2 \times 10^{12}$ and
$10^{13}$, against 10.53 predicted by $N\pi(\sqrt{N})$ and 5.00 by $N$ alone. A linear
extrapolation from $10^{12}$ therefore underestimates the $10^{13}$ time by more than a factor
of two. The rescan is a property of how the tile is sieved, not of the device-resident
design.

\section{Discussion}
\label{sec:discussion}

\subsection{Architectural progression}

The two GoldbachGPU papers represent a sequential elimination of bottlenecks. The first
\cite{llorente2026goldbach} demonstrated that VRAM exhaustion was an architectural artefact
of monolithic prime-table storage, solvable by segmentation at O(1) VRAM cost (14\,MB at default parameters).
That work then exposed a second bottleneck: the host CPU, which had to regenerate and
transfer each segment bitset. The present architecture resolves this by making the GPU
self-sufficient for per-segment sieve generation; the small-prime bitset and base-prime list
are built once on the host at initialisation (\S\ref{sec:gpu_sieve}). Host-device communication per segment is then constant with respect to segment size, for fixed $P_{\text{SMALL}}$ and $P_{\text{BATCH}}$: rather than transferring the entire segment bitset, the host uploads the candidate-prime array ($\approx 628$\,KB at $P_{\text{SMALL}}=10^6$) and downloads a 4-byte reduction integer.

\subsection{Limitations}
\label{sec:limitations}

\paragraph{Multi-GPU efficiency.}
Two-GPU scaling at $10^{12}$ is near-linear, with a speedup of 2.03 (Table~\ref{tab:multi_gpu}).
Four GPUs were not measured. The fixed overheads that limit efficiency, initialisation and the drain of
the final $k$ segments, shrink relative to total time as $N$ grows, the drain's share scaling
as $O(k/S)$ for $S$ segments, so efficiency should stay near unity at larger $N$. This has not
been measured beyond two GPUs.

\paragraph{Sieve cost.}
Each tile rescans the whole base-prime list, so sieve cost grows as $N\pi(\sqrt{N})$ rather
than $N$ (\S\ref{sec:scaling}). At $10^{13}$ the sieve accounts for 87.8\% of kernel time, and
the cost per even number continues to rise with $N$. This bounds how far the present design
can be pushed without changing how the segment bitset is built (\S\ref{sec:scaling}).

\paragraph{Later releases.}
Releases after v2.0.2 (v3.x) change how the segment bitset is built, removing both the per-tile
rescan and the atomic clears introduced by the D1 correction (\S C.2), and replace the Phase~1
kernel. They are outside the scope of this paper, will be described separately, and their
timings are not comparable with those reported here. Every figure in this paper comes from
v2.0.2 (Table~\ref{tab:releases}).

\paragraph{Representational ceiling at $2^{64}$.}
Sieve, indexing and range arithmetic use 64-bit operands, so the framework cannot represent
$n \ge 2^{64} \approx 1.84 \times 10^{19}$; the Miller--Rabin modular multiplication widens to
a 128-bit intermediate within that range (\S\ref{sec:bounds}). Extending past it would require wider integer
types throughout; the twelve-base Miller--Rabin test itself remains deterministic to
$3.18 \times 10^{23}$ \cite{sorenson2017}. The practical ceiling is lower than the
representational one. Two sites are unguarded near $2^{64}$: the segment scheduler can wrap
within $2(k+1)\,\text{SEG\_SIZE}$ of the boundary, about $10^{9}$ at the parameters used here,
and the sieve's alignment computation $(q_{\text{low}} + p - 1)$ overflows when
$q_{\text{low}}$ lies within $P_{\text{SMALL}}$ of it. Neither is reachable at the limits
reported here, and both would need fixing before running near the ceiling.

\paragraph{NVLink vs.\ PCIe topology.}
The lock-free pool uses host-memory atomics, requiring each GPU worker thread to perform a
single atomic operation on system DRAM per segment. On PCIe-only topologies this incurs
one cache-coherency round trip; on NVLink platforms it could be replaced by an on-device
atomic, potentially improving scaling at high GPU counts. This has not been evaluated.

\paragraph{Arbitrary-precision range verification.}

The prior GoldbachGPU framework \cite{llorente2026goldbach} included a \texttt{big\_check} tool verifying individual
even integers beyond the 64-bit range (up to $10^{10\,000}$) using the GNU Multiple
Precision library. Extending this to exhaustive \emph{range} verification is computationally
infeasible: at $d = 100$ digits, a single GMP Miller-Rabin test already costs milliseconds,
and exhaustively testing even a narrow interval containing $\approx 5 \times 10^{99}$ even
integers would require an intractable number of operations. Arbitrary-range
verification at large scales therefore remains an open problem, requiring fundamentally
different techniques such as analytic bounds on the Goldbach comet combined with certified
interval arithmetic.

\subsection{Future Work}
\label{sec:future}

The directions below are stated for the design described here; the per-tile rescan and the
Phase~1 kernel are taken up in later releases (\S\ref{sec:limitations}).

\paragraph{Bitwise bulk-marking kernel.}
The current Phase~1 kernel assigns one GPU thread to each even $n$ in the segment, looping over candidate primes $p$ and performing individual lookups to verify $q = n - p$. Oliveira e Silva's CPU implementation \cite{oliveira2014} takes the opposite approach: it represents the verification array as a bitset and uses 64-bit word-level \texttt{AND}/\texttt{OR} operations to bulk-mark entire ranges of verified integers at once. A GPU kernel adopting this strategy (representing \texttt{d\_verified} as a bitset and replacing per-$n$ loops with warp-wide bitwise marking) could amortise work across 32–64 integers per instruction and substantially increase memory throughput. Such a design would more closely resemble the classical $O(N \log\log N)$ behaviour of
segmented sieves. It addresses Phase~1, which the profile in \S\ref{sec:profiling} puts at
20.3\% of kernel time at $2 \times 10^{12}$ and 11.9\% at $10^{13}$; the segment sieve accounts
for the rest, and the per-tile rescan described in \S\ref{sec:scaling} is the larger target.
The \texttt{d\_verified} clear costs 535\,ms, 0.15\% of kernel time, so the bandwidth saving
from a bitset is small and the added encoding cost may offset it.

\paragraph{Pushing past the 64-bit ceiling.}
Going beyond the representational limit of $2^{64} \approx 1.84 \times 10^{19}$, to $10^{20}$
for example, requires multiword arithmetic throughout the range, indexing and sieving code,
with explicit carry handling. The twelve-base Miller--Rabin test
remains deterministic to $3.18 \times 10^{23}$ \cite{sorenson2017}, so a change of primality
test is not required at that boundary. Baillie-PSW has no known pseudoprime but no proof of
correctness in any range, so it would trade a certificate for speed.

\paragraph{Multi-node distribution via \texttt{--start}.}
The \texttt{--start=N} flag enables disjoint range partitioning across independent compute
nodes. Wrapping this in a trivial MPI or job-scheduler layer would transform the framework
into a distributed verifier, suitable for deployment on HPC clusters without any code
changes to the core CUDA kernels.

\paragraph{CUDA Graphs.}
The three-kernel pipeline (sieve $\to$ Phase 1 $\to$ reduction) is launched sequentially
each segment. Capturing this as a CUDA Graph would amortise kernel-launch overhead ($\approx
5$--$10\,\mu$s per launch) and could improve throughput on Blackwell hardware where kernel
latency is non-negligible relative to segment duration at small $N$.

\section{Conclusion}
\label{sec:conclusion}

We have presented a device-resident architecture for GPU-accelerated verification of
Goldbach's conjecture. By migrating the segmented sieve into GPU shared memory and having each
device claim segments from a lock-free atomic counter, we eliminate the host-device bottleneck
that constrained multi-GPU scaling in prior work.

On identical hardware the architecture is 13.1 times faster than its host-coupled predecessor
at $N = 10^{10}$. It verifies Goldbach's conjecture to $10^{12}$ in 145\,s and to $10^{13}$ in
3246\,s on a single RTX 5090, and to $10^{13}$ in 1577\,s on a second machine with two GPUs,
where scaling at $10^{12}$ is near-linear (speedup 2.03). No counterexample was found. Sieve
cost grows as $N\pi(\sqrt{N})$, which sets the practical limit on this design, and 64-bit
arithmetic sets a representational one at $1.84 \times 10^{19}$.

The framework is open-source, reproducible on commercially available hardware, and deployable across
distributed GPU clusters through its range-partitioning interface.

\section*{Corrigendum}
\label{sec:corrigendum}
\addcontentsline{toc}{section}{Corrigendum}

\noindent\textit{Version~2, 19 September 2026. This version corrects version~1, which remains
available. Changed passages in the body carry a marker of the form [C.2] pointing here;
wording-only corrections (\S C.4) are made in place without a marker. Section and table
numbers are unchanged from version~1. ``Version~1'' and ``version~2'' refer to arXiv versions
of this paper; software releases are written v1.1.0, v2.0.0 and v2.0.2.}

\subsection*{C.1\quad Summary}

The claim of version~1, that every even number up to $10^{13}$ is a sum of two primes, is
confirmed by the corrected release; the version~1 computation itself cannot be relied on. The
claim lies inside the range verified by Oliveira e Silva, Herzog and Pardi
\cite{oliveira2014}, so it was never in mathematical doubt; what this paper contributes is an
independent computation.

Release v2.0.0, the implementation this paper
describes, contained two concurrency defects that could leave a composite number marked prime
(\S C.2). Phase~1 could then accept a partition $n = p + q$ in which $q$ is composite, without
producing a visible error. A third defect, in thread-index arithmetic, is unreachable at the
parameters used here and is recorded for completeness.

Both concurrency defects are corrected in release v2.0.2, which repeats the verification to
$10^{13}$ with zero Phase~2 fallbacks. Each correction is tested: the corrected segment-sieve
kernel passes the differential test that the defective one fails, and the host-side bitset
builder has a regression test that builds with the release (\S\ref{sec:repro}).

All performance figures in version~1 were produced by v2.0.0. Section~\ref{sec:results}
replaces them with measurements of the corrected release, which are higher: 145\,s at $10^{12}$ on a single RTX~5090, against 36.5\,s. Most
of the difference is the cost of the correction. Measured on one machine at identical
parameters, the corrected build takes 2.4, 3.0 and 3.4 times as long as the unfixed build at
$10^{10}$, $10^{11}$ and $10^{12}$, with the remaining 13 to 16\% following from a change of
machine and toolchain. The unfixed timings do not measure a valid computation and appear
nowhere in this paper.

The four-GPU configuration was not re-measured. The figure of $10^{13}$ in 133.5\,s and the
four-GPU parallel efficiency of 98.6\% are withdrawn without replacement. Near-linear scaling
at two GPUs survives the correction (\S\ref{sec:multi_gpu_results}).

\subsection*{C.2\quad The defects}

\paragraph{D1, segment sieve.}
\texttt{tiled\_sieve\_segment\_kernel} cleared bits in a shared-memory bitset with a plain
read-modify-write. Threads of a block sieve distinct primes into the same 64-bit words, so
concurrent clears could be lost, and a lost clear leaves a composite marked prime. Phase~1
reads this bitset for every $q$ above the small-prime bound, which at these limits is almost
every $q$ it tests.

The defect was reported by an independent correspondent, who identified it by reading the
published source. We confirmed it by differential testing of the v2.0.0 kernel, extracted
unchanged, against an independently written CPU segmented sieve over 7 fixed and 75 randomised
ranges covering $12{,}202{,}140$ odd numbers, of which $11{,}005{,}890$ are composite. Across
five runs a mean of $299{,}984$ composites, 2.73\%, were left marked prime. No prime was ever marked composite. Phase~1
examines roughly fourteen candidate values of $q$ per even number before reaching a partition.
The proportion of numbers whose accepted partition could have rested on a false witness is
therefore higher than this rate. That proportion was not measured; the verification has since
been repeated on corrected code. The fix replaces the non-atomic clear with
\texttt{atomicAnd}. Run against the corrected kernel, extracted from v2.0.2 and matched to the
release by checksum, the same harness reports no disagreement in five runs over identical
ranges, while in the same build the defective kernel still fails.

\paragraph{D2, host-side bitset construction.}
\texttt{build\_prime\_bitset} builds the small-prime bitset once on the host, giving each
OpenMP thread a slice of the output. The slices are exclusive in bit index and share 64-bit
words at their boundaries, and \texttt{PrimeBitset::clear} is a non-atomic read-modify-write,
so adjacent threads could lose each other's clears.

This bitset supplies the candidate primes $p$ used by Phase~1, the base primes used by the
segment sieve, and the primality of $q$ up to the small-prime bound. A composite left marked
prime could therefore enter the candidate list and be accepted as $p$. Contamination of the
base-prime list is harmless, since marking multiples of a composite marks nothing that is not
already composite.

With the fix reverted, 23, 33 and 87 of 250 builds at limit $10^6$ were wrong at 4, 8 and 16
threads. Every wrong bit lay within 63 bit positions of a slice boundary, with a measured
maximum of 55, and a limit chosen to make every slice a whole number of words produced no
wrong build at any thread count. The fix rounds each slice up to a whole number of 64-bit
words, with a regression test that builds with the release. The OpenMP thread count of the version~1 runs was not recorded
and the race is nondeterministic, so whether any particular run was affected cannot be
determined. D2 predates v2.0.0 and therefore also affects arXiv:2603.02621
\cite{llorente2026goldbach}, which will be corrected separately.

\paragraph{D3, thread-index overflow.}
\begin{sloppypar}
Every kernel computes its global index as
\texttt{blockIdx.x * blockDim.x + threadIdx.x}. Both operands are \texttt{unsigned int}, so
the product forms in 32 bits and wraps above $2^{32}$ threads. Phase~1 and the counting kernel
wrap identically, so an even number at an offset at or above $2^{32}$ within a segment is
neither verified nor counted, and the program reports success over a range it has not
examined. At a segment size $S$ above $2^{32}$, $S - 2^{32}$ numbers per segment are skipped.
Run at \texttt{--seg-size=4500000000}, where $205{,}032{,}704$ of the segment's
$4{,}500{,}000{,}000$ numbers are unreachable, v2.0.1 printed its success message and reported
zero fallbacks. Every run reported here used \texttt{--seg-size=200000000}, so no figure is
affected. Release v2.0.2 rejects larger segment sizes at startup.
\end{sloppypar}

D1 and D2 are one kind of fault: a bit-packed structure written concurrently by workers whose
partition is exclusive in bit index and not in machine word. D3 is of a different kind, an
index computed in 32 bits. All three can cause the program to report success for values it has
not correctly verified, so a run affected by any of them looks exactly like a correct one. A
verification code therefore needs a differential test of every structure its primality
decisions rest on, checked against an independent implementation and run on the configuration
being reported.

\subsection*{C.3\quad Corrected measurements}

All figures in this version come from release v2.0.2, which is v2.0.0 with the D1 and D2
corrections, the D3 guard, and no other change to the verifier. Parameters throughout:
\texttt{--seg-size=200000000 --p-small=1000000 --batch-size=2000000}. No run triggered a
Phase~2 fallback.

Two machines were used, and no table combines them. \textbf{Machine A}, one RTX~5090,
CUDA~13.3, driver 610.57.01, Ubuntu 26.04 under WSL2, GCC 15.2.0, produced the single-GPU
figures in Table~\ref{tab:algo_speedup} and the profile in Table~\ref{tab:nsys}.
\textbf{Machine B}, two RTX~5090 on an AMD EPYC 9654, CUDA~12.8, driver 575.57.08,
Ubuntu 24.04, GCC 13.3.0, produced Table~\ref{tab:multi_gpu}.

Both binaries on machine~A were built for \texttt{sm\_120}. A default configure of this release
resolves to \texttt{sm\_75} on that machine and runs from JIT-compiled PTX, which does not
reproduce the timings; \S\ref{sec:repro} gives the required flag.

Uncertainties are sample standard deviations with an $n-1$ denominator over the stated number
of runs. Times are the computation time the program reports, which excludes about 0.23\,s of
host initialisation. Command lines, environments and raw output for every figure are in
\texttt{corrigendum-logs/} in the release.

\subsection*{C.4\quad Other corrections}

Four quantitative statements are corrected. \S\ref{sec:results} stated that 78,498 candidate
primes are evaluated per even integer; each thread stops at the first $p$ for which $n - p$
is prime, so the mean is about 14. The same section cited a predicted bound of 4,305 on
the largest $p_{\min}(n)$ up to $10^{13}$, attributed to \cite{oliveira2014, llorente2026goldbach}; it now
cites the largest value recorded in the data of \cite{oliveira2014, oliveiraData}, 3,917.
\S\ref{sec:algo_speedup} reported the speedup over the prior architecture growing from 13.2 at
$10^{9}$ to 45.6 at $10^{10}$ and inferred that the prior architecture was asymptotically
I/O-bound. Both figures are withdrawn. The corrected speedups are 6.0 and 13.1, or 14.9 at both
limits once the setup cost is removed, so they give no support to that inference; the
host-bound diagnosis of the prior architecture rests on the H100/RTX~3090 comparison in
\cite{llorente2026goldbach}, which is unaffected. Table~\ref{tab:nsys} averaged two
\texttt{cudaMemset} operations of different size into one row; they are now listed
separately.

Three descriptions of the implementation are corrected where they bore on the defects or on
their reproduction. \S\ref{sec:bounds} described the GPU as relying on the Miller--Rabin test
for $q$ above the small-prime bound; every $q$ is resolved from one of the two sieved bitsets,
which is where D1 and D2 lay, and the ceiling given as $2^{63.8}$ is $2^{64}$.
\S\ref{sec:phase2} described the Goldbach-comet envelope as guaranteeing a small-prime
partition; it is an empirical regularity, and \S\ref{sec:phase2} now states the $p \le 10^8$
bound on Phase~2. \S\ref{sec:cli} described the start-up VRAM check as preventing
deep-execution OOM; it does not model the CUDA context.

Five further statements were inaccurate and are corrected in place without a marker:
\S\ref{sec:multi_gpu} called the scheduler a work-stealing pool (it is centralised dynamic
self-scheduling on a shared counter); Figures~\ref{fig:simplified_pipeline}
and~\ref{fig:explicit_pipeline} and \S\ref{sec:gpu_sieve} called the no-fallback path
zero-copy and said it bypasses PCIe (no mapped host memory is used, and the 4-byte count
returns over PCIe each segment); \S\ref{sec:bounds} said a failing worker releases its
allocations cleanly (no cleanup is attempted; the process exits); \S\ref{sec:cli} said a
smaller prime batch lets transfer and compute interleave (all work is on one stream, so they
do not); and \S\ref{sec:repro} said all versions, flags and parameters were recorded as
executed (the host CPU, OpenMP thread count and build architecture of the version~1 runs were
not). Two subsection titles change with the text they head: \S\ref{sec:multi_gpu} and
\S\ref{sec:bounds}, the latter to ``Arithmetic Guards and Primality Testing'', since the
guards it describes address overflow and not correctness in general.

\subsection*{C.5\quad Scaling with $N$}

The corrected ladder grows by 10.6, 12.9 and 22.4 between successive decades, because each
tile rescans the whole base-prime list. Version~1 does not discuss scaling with $N$;
\S\ref{sec:scaling} is added and identifies the rescan as the limit of the present design.

The releases referred to above, with their archived DOIs, are listed in
Table~\ref{tab:releases}.

\section*{Software Availability}

The complete source code is available under an open-source licence at
\url{https://github.com/isaac-6/goldbach-gpu}. Table~\ref{tab:releases} lists the releases
referred to in this paper.

\begin{table}[ht]
\centering
\caption{Software releases referred to in this paper.}
\label{tab:releases}
\small
\begin{tabular}{l p{0.47\linewidth} l}
\toprule
Release & Role & Archive \\
\midrule
v1.1.0 & Host-coupled design of \cite{llorente2026goldbach}; source of the comparator in
Table~\ref{tab:algo_speedup} & \href{https://doi.org/10.5281/zenodo.18837081}{\nolinkurl{10.5281/zenodo.18837081}} \\
v2.0.0 & Described in version~1; contains the defects of \S C.2 & \href{https://doi.org/10.5281/zenodo.18879797}{\nolinkurl{10.5281/zenodo.18879797}} \\
v2.0.2 & Corrected release; produces every figure in this version & \href{https://doi.org/10.5281/zenodo.22831383}{\nolinkurl{10.5281/zenodo.22831383}} \\
\bottomrule
\end{tabular}
\end{table}

Release v2.0.2 is v2.0.0 with the corrections of \S C.2 and no other change to the verifier. It
is tagged at commit \texttt{02c6d49}, with raw run logs, environments and command lines under
\texttt{corrigendum-logs/}; the differential-test logs for the corrected kernel follow the tag
on the same branch, at commit \texttt{c3a48fd}. The repository's main branch documents later
releases, whose figures are not those of this paper. Later releases are archived under the concept
DOI \href{https://doi.org/10.5281/zenodo.18786328}{\nolinkurl{10.5281/zenodo.18786328}}.

\section*{Acknowledgements}

The author thanks the developers of CUDA and OpenMP, and the correspondent who reported the
segment-sieve defect described in \S C.2. The author acknowledges the foundational
computational work of Oliveira e Silva, Herzog, and Pardi \cite{oliveira2014}, whose
$4 \times 10^{18}$ record remains the benchmark for Goldbach verification.


\end{document}